\documentclass[aps,twocolumn,showpacs,amsmath,amssymb,pra,reprint]{revtex4-1}

\usepackage{graphicx}%
\usepackage{dcolumn}
\usepackage{bm}
\usepackage{braket}
\begin{document}

\title{Theory for the electron excitation in  dielectrics under an intense circularly polarized laser field}
\author{Tomohito Otobe}
\affiliation{Kansai Photon Science Institute, National Institutes for Quantum and Radiological Science and Technology (QST), Kyoto 619-0215, Japan}
\author{Yasushi Shinohara}
\affiliation{Photon Science Center, School of Engineering, The University of Tokyo, Hongo, Tokyo 113-8654, Japan}
\author{Shunsuke A. Sato}
\affiliation{Graduate School of Pure and Applied Sciences, University of Tsukuba, Tsukuba 305-8571, Japan}
\author{Kazuhiro Yabana}
\affiliation{Center for Computational Sciences, University of Tsukuba, Tsukuba 305-8571, Japan}
\begin{abstract}
 We report  a Keldysh-like model for the electron transition rate in  dielectrics 
under an intense circularly polarized laser.
We assume a parabolic two-band system and the Houston function as the 
time-dependent wave function of the valence and conduction bands.
Our formula reproduces the experimental result for the 
ratio of the excitation rate between linear and circular polarizations for $\alpha$-quartz.
This  formula can be easily introduced into simulations of  
 nanofabrication using an intense circularly polarized laser.
\end{abstract}
\maketitle
\section{Introduction}
 Technical developments in femtosecond laser processing have
made it possible to produces nanoscale laser-induced periodic surface structure (LIPSS), and realize,
non-thermal ablation for sub-wavelength resolution \cite{Hirao96, Mazur08, Hirao97, Corkum06}.

Electron excitation in  dielectrics by an intense laser field is the main process 
in laser-matter interactions.
In particular, for femtosecond lasers, electron excitation by multiphoton ionization and tunnel ionization
are the crucial, because  such nonlinear processes gnerate a controllable free-carrier density and 
confine material change to the focal volume.
Therefore, predicting the electron excitation rate using theoretical models and/or numerical 
simulation is important.

We have been developed a first-principles numerical method to explore 
electron excitation under an intense laser field 
using  time-dependent density functional theory (TDDFT)
\cite{Runge84,Bertsch00,Otobe08,Shino10,yabana12}.
This approach is currently the most reliable and accurate method with  feasible computational cost to simulate electron excitation under intense laser fields. 
However, an analytical model may also be a helpful tool to understand
 some fundamental physical processes in laser processing.
 
 Keldysh proposed a theory for the electron excitation rate under an intense linearly polarized laser field\cite{Keldysh65}. 
His approach is very general and can be used to
describe the photoionization of different objects from single
atoms to crystals \cite{Golin14}. 
Because of its that generality, the Keldysh model
has attracted much attention and  become one of the
standard tools in the theory of laser photoionization.
In particular, for atoms and molecules, the Keldysh-Faisal-Reiss (KFR) theory \cite{Keldysh65, Faisal, Reiss80}
, which is an the implementation of the original Keldysh work, is one of the most important theories in understanding the electron-laser interaction.
 
 Recently,  Temnov \textit{et al} reported that the electron excitation rate induced by a circularly polarized laser
 is twice that induced by a linearly polarized laser at the same laser irradiance \cite{Temnov}.
 Circularly polarized lasers are also important as an ultrafast laser waveguides \cite{Herman06},  
 and in controlling laser-induced nanostructures \cite{Huang08}.
The purpose of this work is to construct an analytical formula
for the transition probability in dielectrics including multiphoton and tunneling processes under a 
circularly polarized laser.
We derive the transition probability in a crystalline solid under a circularly polarized laser
assuming a parabolic two-band system and using the Houston function\cite{Houston}
as the time-dependent wave function.

Jones and Reiss \cite{Reiss77} pioneered work on the electron excitation rate under a circular polarized laser employing the $S$-matrix theory.
Although the Keldysh formula treats the time-dependent wave function of the valence and conduction bands as the Houston function and 
includes only the reduced mass,
  the Jones formula treats only the conduction band as the Houston function (Volkov state) and 
  includes the effective mass of valence and conduction bands independently.
 Therefore, a direct comparison between the Keldysh  and  Jones formulas is not possible.
In the case of atoms, Perelomov {\it et. al.} \cite{Perelomov66} reported the analytical formula for the ionization rate under a circularly polarized laser.
 Because our new formula for circular polarization depends only on the reduced mass, it can be compared directly with the Keldysh formula.
 We also construct a new formula for a linear polarization using a parabolic two-band system.
 The relative ratio of the electron excitation rate between our
  two formulas shows reasonable agreement with the experimental results obtained by Temnov \textit{et al} \cite{Temnov}.

The present article is organized as follows. In section II,
we present our formalism to calculate the transition probability per unit time in a crystalline solid. 
In section III, we describe our results for  $\alpha$-quartz with  linearly and circularly polarized laser fields. 
A summary is presented in section IV.

\section{Formalism}
\subsection{Houston function}
The static Schr\"odinger equation for a spatially periodic  system in atomic units is
\begin{equation}
\epsilon_{n,\vec{k}}^G u_{n,\vec{k}}^G(\vec{r})=\left[\frac{1}{2m}(\vec{p}+\vec{k})^2+V(\vec{r})\right]u_{n,\vec{k}}^G(\vec{r}),
\end{equation}
where $\vec{p}$ is the momentum operator,
$\vec{k}$ is the Bloch wave vector, $V'(\vec{r})$ is the spatially periodic potential, and $\varepsilon_{n,\vec{k}}^G$ and  $u_{n,\vec{k}}^G(\vec{r})$
are the energy  and wave function, respectively,  of the $n$-th band for the Bloch wave vector $\vec{k}$. 
$u_{n,\vec{k}}^G(\vec{r})$ satisfies the periodic boundary condition, $u_{n,\vec{k}}^G(\vec{r}+\vec{R})=u_{n,\vec{k}}^G(\vec{r})$.
The time-dependent Schr\"odinger equation under a time-dependent vector potential is described as
\begin{eqnarray}
\label{eq:TDSE}
&i&\frac{\partial u_{n,\vec{k}}(\vec{r},t)}{\partial t}= \nonumber\\
&\bigg[&\frac{1}{2m}(\vec{p}+\vec{k}+\frac{e}{c}\vec{A}(t))^2+V(\vec{r})\bigg]u_{n,\vec{k}}(\vec{r},t).
\end{eqnarray}
Here $\vec{A}(t)$ is the vector potential of the applied laser field.
Now, we assume the Houston function,
\begin{equation}
w_{n,\vec{k}}(\vec{r},t)=u_{n,\vec{k}+\frac{e}{c}\vec{A}(t)}^G(\vec{r})\exp\left[-i \int^t \varepsilon_{n,\vec{k}+\frac{e}{c}\vec{A}(t')}^G dt'\right].
\end{equation}
The time evolution of the Houston function is described as,
\begin{eqnarray}
\label{eq:Houston_TD}
\frac{\partial w_{n,\vec{k}}(\vec{r},t)}{\partial t}&=&\frac{e}{c}\frac{d\vec{A}(t)}{dt}\frac{\partial u_{n,\vec{k}}^G}{\partial \vec{k}}| _{\vec{k}+\frac{e}{ c}\vec{A}(t)} e^{-i\int^t \varepsilon_{n,\vec{k}+\frac{e}{c}\vec{A}(t')}^G dt'}\nonumber\\
&+&\varepsilon_{n,\vec{k}+\frac{e}{c}\vec{A}(t)}^G u_{n,\vec{k}+\frac{e}{c}\vec{A}(t)}^G(\vec{r}).
\end{eqnarray}
We assume that the time-dependent wave fuction $u_{n,\vec{k}}(\vec{r},t)$ can be expanded by the Houston function,
\begin{equation}
u_{n,\vec{k}}(\vec{r},t)=\sum_{n'}C^{\vec{k}}_{nn'}(t)w_{n,\vec{k}}(\vec{r},t).
\end{equation}
The time evolution of the coefficient $C^{\vec{k}}_{nn'}(t)$ can be expressed by the simple form,
\begin{eqnarray}
&&\frac{\partial C^{\vec{k}}_{nn'}(t)}{\partial t}=\frac{e}{mc}\frac{d \vec{A}(t)}{dt}\frac{\left< u_{n',\vec{k}+\frac{e}{c}\vec{A}(t)}^G\Big|\vec{p}\Big|u_{n,\vec{k}+\frac{e}{c}\vec{A}(t)}^G\right>}{\varepsilon_{n',\vec{k}+\frac{e}{c}\vec{A}(t')}^G-\varepsilon_{n,\vec{k}+\frac{e}{c}\vec{A}(t')}^G}\nonumber\\
&\times&\exp\left[-i \int^{t} dt' \left(\varepsilon_{n',\vec{k}+\frac{e}{c}\vec{A}(t')}^G-\varepsilon_{n,\vec{k}+\frac{e}{c}\vec{A}(t')}^G \right)\right]
\end{eqnarray}
The transition from $w_{n,\vec{k}}(\vec{r},t)$ to $w_{n',\vec{k}}(\vec{r},t)$  in an arbitrary time interval [-T,T] has the following form
\begin{eqnarray}
\label{eq:Cnn}
\tilde{C}_{nn'}^{\vec{k}}&=&-\frac{ie}{mc}\int_{-T}^{T} dt \vec{P}^{\vec{k}}_{n'n}\cdot \vec{A}(t) \nonumber\\
&\times&\exp \left[-i \int^{t} dt' \left(\varepsilon_{n',\vec{k}+\frac{e}{c}\vec{A}(t')}^G-\varepsilon_{n,\vec{k}+\frac{e}{c}\vec{A}(t')}^G \right) \right]
\end{eqnarray}
where $ \vec{P}^{\vec{k}}_{n'n}$ is the transition momentum matrix,
\begin{equation}
\vec{P}^{\vec{k}}_{n'n}=\langle u^G_{n',\vec{k}}(\vec{r})|\vec{p} |u^G_{n,\vec{k}}(\vec{r}) \rangle.
\end{equation}

\subsection{Parabolic two-band system}
 The Keldysh formula assumes the band structure is,
 \begin{equation}
 \varepsilon_{c\vec{k}}-\varepsilon_{v\vec{k}}=B_g\sqrt{1+\frac{\vec{k}^2}{\mu B_g}}.
 \end{equation}
In contrast, we used a parabolic two-band system: i.e.
\begin{equation}
\varepsilon^G_{c,\vec{k}}-\varepsilon^G_{v,\vec{k}}=B_g+\frac{\vec{k}^2}{2\mu},
\end{equation}
where index $c$ ($v$) represents the conduction (valence) band, $B_g$ is the band gap, and $\mu$ is the reduced mass.

\subsubsection{Circular polarization}
We assumed a circularly polarized laser field,
\begin{equation}
\vec{A}(t)=A_0(\hat{x}\cos\omega t + \hat{y}\sin\omega t),
\end{equation}
where $\hat{x}$ and $\hat{y}$ are the unit vectors along the $x$- and $y$-directions, respectively.
We assumed that the propagation direction of the light was along the $z$-axis.
The coefficient $C_{nn'}^{\vec{k}}$ in Eq.~(\ref{eq:Cnn})  for the two-band system ($C_{cv}^{\vec{k}}$ ) can be written as
\begin{eqnarray}
\label{eq:Ccv}
&&C_{cv}^{\vec{k}}=-\frac{ieA_0}{2mc}\int_{-T}^{T} dt\left(M_{vc\vec{k}}^- e^{i\omega t}+M_{vc\vec{k}}^+ e^{-i\omega t}\right)\nonumber\\
&\times&\exp \left[-i \int^{t} dt' \left(\varepsilon_{c,\vec{k}+\frac{e}{c}\vec{A}(t')}^G-\varepsilon_{v,\vec{k}+\frac{e}{c}\vec{A}(t')}^G \right) \right],
\end{eqnarray}
where  
\begin{equation}
M_{vc\vec{k}}^{\pm}=P_{vc,x}^{\vec{k}}\pm iP_{vc,y}^{\vec{k}}.
\end{equation}
The exponential part of Eq.~(\ref{eq:Ccv}) can be expanded by  Bessel functions
\begin{eqnarray}
&&\exp \left[-i \int^{t} dt' \left(\varepsilon_{c,\vec{k}+\frac{e}{c}\vec{A}(t')}^G-\varepsilon_{v,\vec{k}+\frac{e}{c}\vec{A}(t')}^G \right) \right] \nonumber\\
&=&\exp\left[-i \left(B_g+U_c+\frac{k^2}{2\mu} \right)t\right]\sum_l J_l(\eta) e^{il(\omega t-\phi)},
\end{eqnarray}
where $\phi$ is the angle between $(k_x,k_y,0)$ and the $x$-axis, 
 $\eta=ekA_0\sin\theta/\mu\omega c$, and 
 $U_c$ is the averaged kinetic energy of the charged particle in the circular polarized laser, 
$U_c=e^2A_0^2/2\mu c^2$. 
The band gap is blue shifted by $U_c$ \cite{Jauho96}, and 
$\theta$ is the angle between the $z$-axis and $\vec{k}$.

The time-averaged transition probability per unit time and space, $\tau_k$, is found from
\begin{eqnarray}
\tau_k&=&\lim_{T\rightarrow \infty} \frac{|\tilde{C}^{\vec{k}}_{cv}|^2}{2T}=\frac{e^2\pi A_0^2 |P_{vc}|^2}{2m^2c^2}\nonumber\\
&\times&\sum_l\Big( J_{l-1}^2(\eta)+J_{l+1}^2(\eta)\Big)\delta(\xi_k),
\end{eqnarray}
where $\xi_k=(B_g+U_c+k^2/2\mu +l\omega)$. 
In this step, we assume that $\vec{P}_{vc}^{\vec{k}}$ do not depend on $\vec{k}$. 

Gertsvelf {\it et al} have reported that orientation dependence on the electron excitation rate is few ten's \%  even for $\alpha$-quartz \cite{Gertsvelf08}.
This fluctuation is minor effect for our purpose in this work, estimation in order and/or factor accuracy.
In the case of the interaction between a circularly polarized laser and solid, one may consider that the 
angular momentum conservation defines the selection rule for a transition. 
While the angular momentum transfer also depends on lattice structure and dynamics \cite{Simon68}, 
we only focus on the electronic response in this theory.

The total transition probability induced by the laser field, $W$, is found to be:
\begin{eqnarray}
\label{Rate_C}
W&=& \frac{e^2 A_0^2 |P_{vc}|^2\mu^{3/2}}{\sqrt{2}\pi m^2c^2}\sum_{l=l_0}^{\infty}\int d\theta \sin\theta \nonumber\\
&\times& \Big( J^2_{l-1}(\eta')+J^2_{l+1}(\eta')\Big)\sqrt{\zeta_{l}},
\end{eqnarray}
where, 
 \begin{equation}
 \eta'=\frac{eA_0\sqrt{2\zeta_{l}}\sin\theta}{\sqrt{\mu}\omega c},
 \end{equation}
 and 
 \begin{equation}
 \zeta_{l}=  l \omega-(B_g+U_c).
 \end{equation}
 In Eq.~(\ref{Rate_C}), we changed the definition of $l$ to $-l$.
The lowest order of $l=l_0$ is the positive minimum value for $\zeta_{l_0}>0$.
Here,  $\theta$ is the angle between the propagation direction of the laser ($z$-axis) and  $\vec{k}$.
The integration about the $|\vec{k}|$ is replaced by the summation about $l$ because of  the $\delta$-function in Eq.~(\ref{Rate_C}).

In the low intensity limit, the dominant term in $W$ is $J_{l_0-1}$ which has a $A_0^{2(l_0-1)}$
dependence.
Because this coefficient includes $A_0^2$, the intensity dependence has the usual 
multiphoton absorption behavior of $W\propto I^{l_0}$. 


\subsubsection{Linear polarization}
To compare  linear and the circular polarizations, 
here we revisit the excitation rate under a linearly polarized laser.
We assumed a linearly polarized continuous wave field,
\begin{eqnarray}
\vec{A}(t)&=&A_0\hat{u}\cos\omega t,\\
\hat{u}&=&(0,0,1).
\end{eqnarray}
Following a similar procedure  to that used for circular polarization, the total transition probability, $W_L$, is found from
\begin{eqnarray}
W_L&=&
 \frac{e^2 A_0^2 |P^{\vec{k}}_{vc}|^2\mu^{3/2}}{2\sqrt{2}\pi m^2c^2} \int d\theta'\sin\theta'\nonumber\\
&\times&\sum_{l=l_0}^{\infty} ( J_{l-1}(\alpha',\beta)+J_{l+1}(\alpha',\beta)\Big)^2\sqrt{\kappa_l},
\end{eqnarray}
where 
$\kappa_l=l \omega-(B_g+U_p)$,
$\theta'$ is the angle between the polarization direction and  $\vec{k}$,
and $l_0$ is the maximum integer $l$ so that $\kappa_l > 0$.
$J_l(\alpha,\beta)$ is the generalized Bessel function \cite{Reiss03} and 
$U_p=e^2A_0^2/4\mu c$ is the ponderomotive energy.
Here, $\alpha'$ and $\beta$ are defined as
\begin{equation}
\alpha'= \frac{eA_0\sqrt{2\kappa_l}\cos\theta'}{\sqrt{\mu}\omega c},
\end{equation}
and
\begin{equation}
\beta= \frac{e^2A_0^2}{8\mu\omega c^2}.
\end{equation}
\section{Application to $\alpha$-quartz}
 \begin{figure}
\includegraphics[width=90mm]{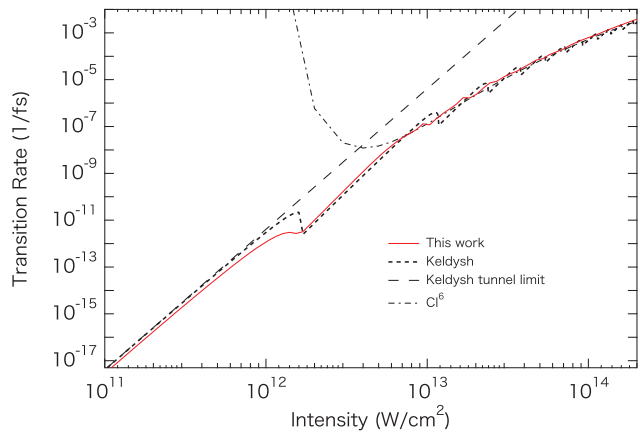}
  \caption{\label{fig:OK}
  Transition probability  as a function of  
laser  intensity  for linearly polarized 800-nm light in  $\alpha$-SiO$_2$.
The red solid line represents the excitation rate determined by our formalism, 
the blue dashed line represents the excitation rate 
based on the full expression of the Keldysh theory, the green dotted line
represents the tunneling limit of the Keldysh theory, and the black dot-dashed line represents 
the simple six-photon process. }
      \end{figure}
\begin{figure}
\includegraphics[width=90mm]{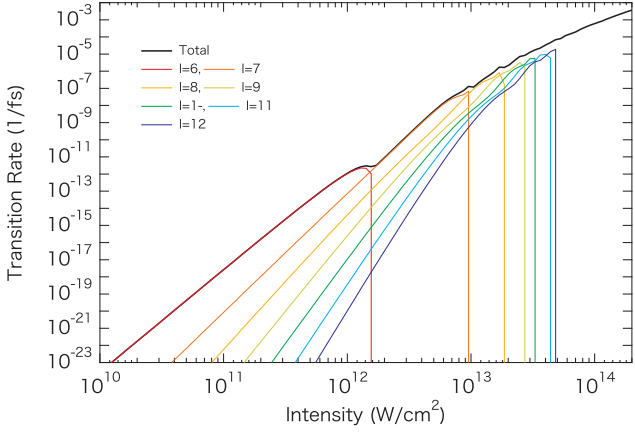}
\caption{\label{fig:PN} 
Total transition probability for linear polarization as a function of  laser intensity. 
The solid red curve is the same as that in  Fig.\ref{fig:OK}.
The curves labeled $l=6,7,\cdots,12$ give the separate contributions of each order.}
\end{figure}

\begin{figure}
\includegraphics[width=90mm]{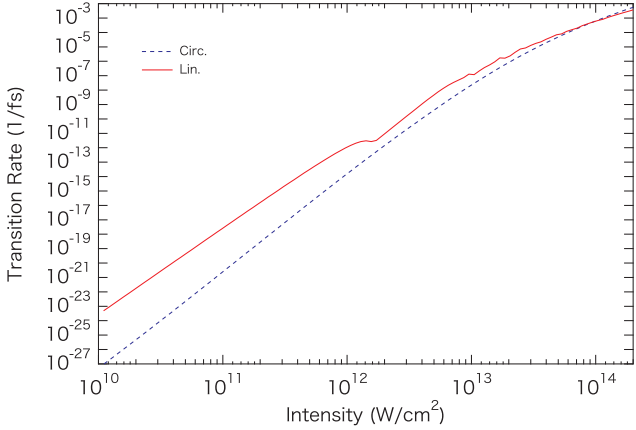}
\caption{\label{fig:LC}
Transition probability as a function of  electric field intensity for  linearly polarized (red solid line) and  
circularly polarized (blue dashed line) light. }
\end{figure}

\begin{figure}
\includegraphics[width=90mm]{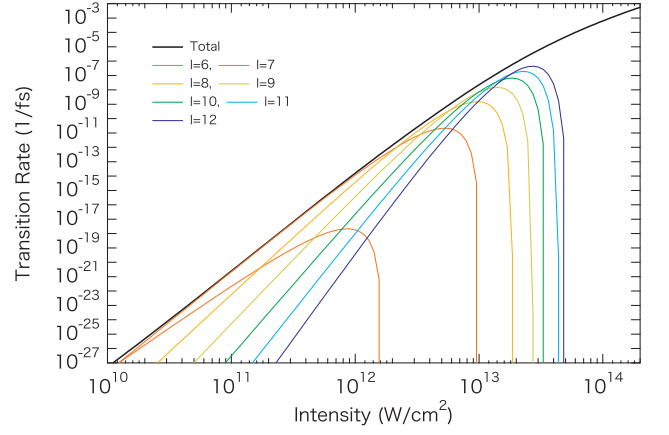}
\caption{\label{fig:PNCir}  
Total transition probability for circular polarization as a function of electric field intensity. 
The solid red curve is the same as the blue dashed curve in Fig.\ref{fig:LC}.
The curve labels $l=6,7,\cdots,12$ give the separate contribution of each order.}
\end{figure}

\begin{figure}
\includegraphics[width=90mm]{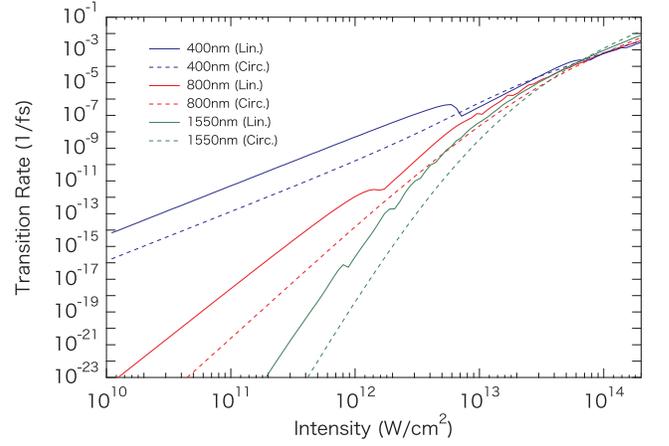}
\caption{\label{fig:WL} 
Wavelength dependence of the total transition probability for circular and linear polarizations as a function of laser intensity.
Solid curves represent linear polarization, and dashed lines represent circular polarization.}
\end{figure}
\subsection{Linear polarization}

 $\alpha$-Quartz is a typical dielectric used in  non-linear laser-matter interaction studies, and we   
selected it here as an example with which to illustrate the application of our developed formalism.
The transition probability ($W_L$) of $\alpha$-SiO$_2$ by  linearly polarized 800nm light is shown in Fig.~\ref{fig:OK} by a red solid line.
We assumed a band gap ($B_g$) of 9 eV and  reduced mass of 0.30$m$ \cite{Brar96}.
For the momentum matrix element $P^{\vec{k}}_{cv}$  we applied the Kane two-band model \cite{Kane} giving,
\begin{equation}
|P^{\vec{k}}_{cv}|^2\sim \frac{m^2}{4}\frac{B_g}{\mu}.
\end{equation} 
In the case of $\alpha$-quartz, we assumed $|P_{cv}|^2\sim 0.28$ a. u. for $B_g=9$ eV and $\mu=0.30m$. 
$T$ calculated by the full expression of the conventional  Keldysh formula (dashed line) 
and  tunneling limit (dot-dashed line) are also shown in Fig.~\ref{fig:OK} for 
 comparison.
Our formalism shows excellent agreement with the Keldysh theory. 
This result indicates that our formula includes the multiphoton and tunneling processes as the Keldysh formula does.

At lower intensity, the transition probability is expected to depend on the laser intensity $I$ as $W_L\propto CI^l$, with $l=6$ in the multiphoton absorption
picture. We show a curve of this dependence in Fig.~\ref{fig:OK} with the green dotted line.
Our result maintains this picture up to $5\times10^{11}$ W/cm$^2$.

Because the ponderomotive energy $U_p$ widens the band gap as the laser intensity increases, the contribution of each $l$-photon process also changes.
Figure \ref{fig:PN} presents the contribution of each $l$-photon process to $T$ as a function of the maximum electric field intensity.
At around $1\times10^{11}$ W/cm$^2$, the probability of the six-photon absorption falls to zero, because the effective band gap ( $B_g+U_p$) 
is larger than 6$\hbar\omega$ at this point, and the seven-photon process becomes dominant.
Simultaneously, the change of the order of the multiphoton process induces a small jump in $T$ which is also seen in the Keldysh formula and 
 its extended model developed by Gruzdev\cite{Gruzdev07}.
 This abrupt change of the $l$-photon process indicates that the averaged kinetic energy $U_p$ defines the effective band gap which is similar to
 the blue shift of the band gap in the dynamical Franz-Keldysh effect \cite{Jauho96,Nordstorm98}.
 However, the critical intensity for higher multi-photonic processes is slightly different from that determined by 
 the Keldysh formula which  may originated from the definition of the band structure.

In a higher intensity region, over $1\times10^{13}$ W/cm$^2$, the contributions from some $l$-photon processes are comparable. 
 In this intensity region, the tunneling model is valid because the Keldysh parameter, $\gamma=c\sqrt{\mu B_g}/eA_0$, is less than one.
Therefore, this result shows that the tunneling process can be interpreted as the summation of  many multi-photon processes.

\subsection{Circular polarization}

The transition rate induced by circularly polarized 800-nm light as a function of laser intensity 
is illustrated  in Fig.\ref{fig:LC} as a blue dashed line. 
The transition rate induced by linearly polarized 800-nm light is also shown as a solid red line.
The power law of the transition rate induced by the circularly polarized light is slightly different from that for  
 linearly polarized light and higher than that for linearly polarized light of higher intensity.
 Temnov \textit{et al} \cite{Temnov} reported the ratio of the ionization rate around an intensity of $1\times 10^{13}$ W/cm$^2$.
 From the relation between the intensity in vacuum ($I_v$) and in media ($I_m$), $I_m=\varepsilon^{1/2} I_v$ where $\varepsilon$
  is the dielectric constant, 
$1\times 10^{13}$ W/cm$^2$  corresponds to $4\times 10^{12}$ W/cm$^2$ in Fig. \ref{fig:LC}.
 The experimental value for the excitation rate ratio $W/W^l$ is  0.3.
 Our result gives a ratio of  $0.1\sim0.2$ at $2\times 10^{12}$ $\sim$ $1\times 10^{13}$ W/cm$^2$ which is in  reasonable agreement with the experimental value.
 
Figure \ref{fig:PNCir} shows the contribution of each $l$-photon process to $W$ as a function of the maximum electric field intensity.
The change of the multiphoton process is not abrupt like in the case with linear polarization (Fig. \ref{fig:PN}) and the contribution
 of each photon processes is more important at lower intensity.
 In the circular polarization, the drop off of each $l$-photon process is more moderate compared with that induced by linearly polarized laser.
 This qualitative difference origineates from the parameter dependence of the generalized and normal Bessel functions.

The contribution of each photon process is also different, because
  the kinetic energy of a charged particle under circularly polarized laser changes from $U_p$ to $U_c$, which effectivly shifts the optical band gap.
Because  $U_c$ is larger than  $U_p$ by a factor of $2$, which corresponds to a factor of $\sqrt{2}$ in the field intensity, 
 $B_g+U_c$ exceeds $6\hbar\omega$ at lower field intensity when polarization is circular rather than linear.

Laser wavelengths other than of 800 nm are also used for laser processing.
For example, 400 nm (3.1 eV), which is the second harmonic of 800-nm, is useful to lower the order of multiphoton processes,
 and 1550 nm (0.8 eV) is the typical wavelength of fiber lasers.
 Figure \ref{fig:WL} depicts the wavelength dependence of the total transition rates under linear (solid) and circular (dashed) polarizations.
 
 For all wavelengths, above $5\times10^{13}$ W/cm$^2$, the transition rate induced by circular polarization is comparable to that induced by linear polarization.
 In contrast, at lower intensity, the ratio between circular and linear polarizations becomes large as the wavelength increases: i.e., as 
 photon energy decreases.

\section{Summary}
We  extended the Keldysh-type formula for the solid state under an intense circularly polarized laser 
assuming the Houston function for the valence and conduction bands.
Because our formula depends only on the reduced mass, it can be directly compared with the Keldysh formula.  
Our simple formula describes electron excitation rate,  
reproduces the Keldysh formula with excellent agreement for $\alpha$-quartz, and makes it possible 
to separate the contribution of each $l$-photon process  with linear or circular polarization. 
The transition rate ratio between linear and circular polarizations determined using our formula shows 
reasonable agreement with experimental results.
\section*{Acknowledgement}
This work was supported by a JSPS KAKENHI (Grants No. 21740303 and No. 15H03674). 
Numerical calculations were performed on the supercomputer PRIMERGY BX900 at 
the Japan Atomic Energy Agency (JAEA).

\end{document}